\documentclass{ws-procs10x7}
\usepackage{balance}

\newcommand{\beq}{\begin{eqnarray}}
\newcommand{\eeq}{\end{eqnarray}}
\newcommand{\be}{\begin{equation}}
\newcommand{\ee}{\end{equation}}

\def\fun#1#2{\lower3.6pt\vbox{\baselineskip0pt\lineskip.9pt
\ialign{$\mathsurround=0pt#1\hfil ##\hfil$\crcr#2\crcr\sim\crcr}}}

\newcommand{{\SD}}{\rm SD}

\newcommand{\ver}{\mbox{\boldmath${\rm r}$}}

\newcommand{{\tr}}{{\rm tr}}

\newcommand{\lan}{\langle}
\newcommand{\ran}{\rangle}

\newcolumntype{d}[1]{D{.}{.}{#1}}

\makeindex
\begin{document}

\title{Debye screening in the hot non-Abelian plasma}

\author{N. O. AGASIAN$^*$ \and YU. A. SIMONOV}

\address{Institute of Theoretical and Experimental Physics,
Moscow, 117218, Russia
\\$^*$E-mail: agasian@itep.ru}


\twocolumn[
\maketitle
\abstract{The  Debye mass $m_D$ is computed
nonperturbatively in the deconfined phase of QCD, where
chromomagnetic confinement is known to be present. The latter
defines $m_D$   to be $m_D=c_D\sqrt{\sigma_s}$, where $c_D \cong 2.06$ and $\sigma_s=\sigma_s(T)$ is the spatial string tension. The
resulting magnitude of $m_D(T)$ and temperature dependence are in
good agreement with  lattice calculations.}
]

\section{Introduction}
The screening of electric fields in QCD was originally considered in
analogy to QED plasma, where  the Debye screening mass was well
understood \cite{1sc}, and the perturbative leading order (LO) result for QCD was
obtained long ago \cite{2sc}, $m_D^{\rm(LO)} =\left( \frac{N_c}{3}
+ \frac{N_f}{6}\right)^{1/2} gT$.
Another difficulty of the perturbative approach is that
the gauge-invariant definition of the one-gluon Debye mass is not available.
Recently the calculations of $m_D(T)$  was computed on the
lattice for $N_f=0,2$ \cite{13sc,14sc,15sc} using the free-energy asymptotics.

The purpose of our talk is to provide a gauge-invariant and
a nonperturbative method, which allows to obtain Debye masses in a
rather simple analytic calculational scheme.
In what follows we use the basically nonperturbative
approach of Field Correlator Method (FCM) \cite{16sc,20sc}
and Background Perturbation Theory (BPTh) for nonzero $T$
\cite{23sc,24sc,25sc} to calculate $m_D(T)$ in a series, where the
first and dominant term is purely nonperturbative \cite{agsim_plb}
\be
m_D(T)=M_0(T) + {\rm BPTh~ series}.
\label{2sc}
\ee
Here $M_0(T)$ is the gluelump mass due to chromomagnetic
confinement in  3d,  which is computed \cite{agsim_plb} to be $M_0(T) = c_D \sqrt{\sigma_s(T)}$,
with $\sigma_s(T)$ being the spatial string tension and $c_D \approx 2$ for $N_c=3$.
The latter is simply expressed in FCM through chromomagnetic correlator \cite{19sc,22sc},
and can be found  from
the  lattice data \cite{38sc,26sc}. Therefore  $M_0(T)$ is
predicted for all $T$ and can be compared with lattice data.

\section{Background Perturbation Theory for the thermal Wilson loop}

One starts with the correlator of Polyakov loops $\lan L(0) L^+ (\ver)\ran$

$$
\lan L(0) L^+ (\ver)\ran = \frac{1}{N^2_c} \exp \left( - \frac{ {F_1(r, T)}}{T} \right)
$$
\be
+ \frac{N^2_c-1}{N^2_c} \exp \left( - \frac{F_8(r, T)}{T}  \right).
\label{4sc}
\nonumber
\ee
The first term
contributing to the free energy  $F_1$ of the static $Q\bar Q$-pair in the singlet color state,
the second to the octet free energy $F_8$.

A convenient way to define Debye mass was suggested in \cite{13sc,14sc}
\be
F_1(r,T) -F_1(\infty, T)
\approx -\frac43 \frac{\alpha_s(T)}{r} e^{-m_D(T)r}
\nonumber
\ee
Accordingly one ends
for $F_1$ with the thermal Wilson loop of time extension $\beta =1/T$ and
space extension $r$,
$$
\exp \left( -\beta F_1(r, \beta)\right) = \lan W (r, \beta)\ran
$$
\be
= \frac{1}{N_c}
\lan \tr P \exp (ig \int_C A_\mu dz_\mu) \ran.
\label{5sc}
\ee
Eq.(\ref{5sc}) is the basis of our approach. In what follows we
shall calculate however not $F_1$, which contains all
tower of excited states over the  ground state of heavy quarks
$Q\bar Q$, but rather the static potential $V_1(r,T)$,
corresponding to this ground state. {As a result the Debye mass $m_D$
is calculated gauge-invariantly in terms of the thermal Wilson loop.

Separating, as in BPTh  the field $A_\mu$ into NP
background $B_\mu$ and valence gluon field $a_\mu$
 \be
 A_\mu=B_\mu +a_\mu \label{16sc}
 \ee
 one can assign gauge transformations as follows
 \be B_\mu\to U^+ (B_\mu +\frac{i}{g} \partial_\mu) U, ~~ a_\mu
 \to U^+ a_\mu U.
 \label{17sc}
 \ee

As a next step one inserts (\ref{16sc}) into (\ref{5sc}) and
expands in powers of $ga_\mu$, which gives an expansion for the Wilson loop
$$
\lan W(r, \beta) \ran =
\lan W^{(0)} (r, \beta) \ran _B+
$$
\be
+ {\lan W^{(2)} (r, \beta) \ran_{B,a}} +\ldots,
\label{18sc}
\ee
and $\lan W^{(2)} \ran$ can be written as
$$
\lan W^{(2)}\ran_{B,a} = \frac{(ig)^2}{N_c} \int dx_\mu dy_\nu
$$
\be
\times \lan
\tr P \Phi (\prod_{xy}) \lan a_\mu (x) a_\nu(y)\ran_a \Phi
(\coprod^{xy}) \ran_B .
\label{19sc}
\ee
Here {$\Phi (\prod)$} and {$\Phi(\coprod)$}  are parallel transporters
along the pieces of the original Wilson loop $W(r, \beta)$, which
result from the dissection of the Wilson loop at points $x$ and
$y$. Thus the Wilson loop $W^{(2)} (r, \beta)$ is the standard loop
$W^{(0)} (r, \beta)$   augmented by the adjoint line connecting points
$x$ and $y$. It is easy to see using (\ref{17sc}), that this
construction is gauge invariant.

For OGE propagator one can write the path integral Fock-Feynman-Schwinger (FFS)
representation  for nonzero $T$
$$
G_{\mu \nu} (x, y)= \lan a_\mu (x) a_\nu (y)\ran_a
$$
$$
=\int^\infty_0 ds \int (D^4z)^w_{xy} \exp (-K) {\Phi^{adj} (C_{xy})}
$$
\be
\times \left(P_F \exp (2ig \int^s_0 F_{\sigma \rho}(z(\tau)) d \tau)\right)_{\mu \nu},
\label{20sc}
\ee
$$\Phi^{adj} (C_{xy})= P\exp (ig \int_{C_{xy}} B_\mu dz_\mu)$$
where the open contour
$C_{xy}$ runs along the integration path in (\ref{20sc}) from the
point $x$ to the point $y$
and $K=\frac14 \int^s_0 (\dot z_\mu)^2 d\tau$.
The $(D^4z)_{xy}^w$ is a path integration measure with boundary conditions
$z_\mu(\tau=0) = x_\mu$ and $z_\mu(\tau=s) =y_\mu$
and with all possible windings in the Euclidean temporal direction
(this is marked by the superscript $w$).

We must now average over $B_\mu$ the geometrical construction
obtained by inserting (\ref{20sc}) into (\ref{19sc}), i.e.
\be
\lan \Phi (\prod_{xy})  \Phi^{adj} (C_{xy} ) \Phi
(\coprod^{xy})\ran_B\equiv \lan W_{xy} (r,\beta)
\ran_B.
\label{21sc}
\ee
For $T>T_c$ one can apply  the
nonabelian Stokes theorem, which yields the area law \cite{16sc}
for distances $r \gg \lambda_g$, $\lambda_g-$ gluon correlation length, $\lambda_g \sim 0.2$ fm
\be
\lan W_{xy}(r,\beta)\ran_B = \exp ( -\sigma_{adj}^H S_{gl}^H),
\label{22sc}
\ee
where $S^{H}_{gl}$ is the area of the space-like projection of gluon-deformed piece of
surface $S_{gl}$, and $\sigma_{adj}^H$ is the adjoint spatial string tension,
$\sigma_{adj}^H = (9/4) \sigma_s$ for $SU(3)$. $ \sigma_s$
is the fundamental spatial string tension.

 As the result one obtains exactly the form containing the gluelump Green's function
$$
\lan W^{(2)}\ran_{B,a} ={(ig)^2} C_2(f)
$$
\be
\times \int^\beta_0 dx_4 \int^\beta_0 dy_4 G_{44}(r, t_4),
 \label{13sc}
 \ee
where  $t_4\equiv x_4-y_4$, $C_2(f)=(N_c^2-1)/2N_c$ and $G_{44} (r, t_4)$ is
$$
G_{44}(r, t_4) =\int^\infty_0 ds \int (D^4z)_{xy}^w
$$
\be
\times \exp (-K) \exp(-\sigma_{adj}^H S_{gl}^H).
\label{14sc}
\ee

Thus the gluon Green's function in the confined phase becomes
a gluelump Green's function, where the adjoint
source trajectory is the projection of the gluon trajectory on the
Wilson loop plane. Gluelump is the system when a gluon is connected by the string to the Wilson loop
plane. String is made of chromomagnetic field only.
Thus the problem reduces to the calculation of the gluelump Green's function
\be
 G(x,y) = \lan x|\sum_n
\exp({-H_n r})|y\ran.
\nonumber
\ee
The lowest eigenvalue of $M_0$ Hamiltonian yields the static $Q\bar Q$ potential
\be
V_1^{(1)} (r,T) =-\frac{N_c^2-1}{2N_c} \frac{\alpha_s}{r} e^{-M_0 r}.
\nonumber
\ee
As a result one finds \cite{agsim_plb} $m_D=M_0$
 \be
m_D=(2.82 - 5.08 \alpha_s^{\rm eff}) \sqrt{\sigma_s} \approx 2.06  \sqrt{\sigma_s}
\nonumber
 \ee
This is the central result of our talk.
Note, that no free parameters are used, $\alpha_s^{\rm eff}$ is taken from glueball
spectrum and BFKL.

\section{Numerical results}

One can now compare our prediction for $m_D(T) =c_D \sqrt{\sigma_s(T)}$
with the latest lattice data \cite{15sc}.
At temperatures $T>T_{dr} \approx (1.2 \div 1.5) T_c$, (where $T_{dr}$
is the temperature of dimensionally reduction and
physical justification for
resorting to dimensionally reduced regime at $T\simeq\sqrt{\sigma_s(T)}$
was given in \cite{22sc,Agasian:1997wv})
the spatial string tension is chosen in the form \cite{38sc,26sc}
\be
\sqrt{\sigma_s(T)}= c_\sigma g^2 (T) T,
\label{sigmas}
\ee
with the two-loop expression for $g^2(T)$
\be
\label{coupl}
g^{-2}(t) =2 b_0 \ln \frac{t}{L_\sigma}+
\frac{b_1}{b_0}\ln \left(2\ln \frac{t}{L_\sigma}\right),
\ee
where $t\equiv T/T_c$.

\begin{figure}
\begin{center}
\includegraphics[width=185pt,keepaspectratio=true]{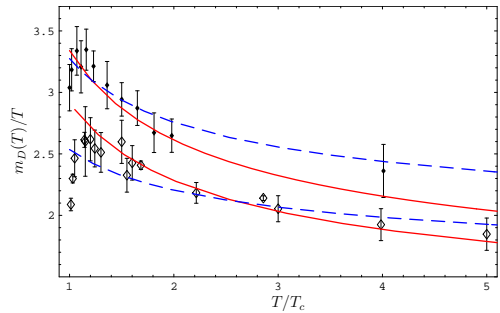}
\end{center}
\caption{Chromoelectric Debye mass $m_D/T$ for 2-flavor
QCD (upper lines) and quenched ($N_f=0$) QCD (lower lines) versus
$T/T_c$. Solid lines are calculated using $m_D(T) =2.09
\sqrt{\sigma_s(T)}$, where $\sqrt{\sigma_s(T)}$ corresponds to
Eq.(\ref{sigmas}) with $N_f=2$ for upper solid line and $N_f=0$ for
lower solid line. Dashed lines are calculated using Eq.(\ref{37sc}),
$N_f=2$--upper line, $N_f=0$--lower line. The lattice data are from
O.Kaczmarek and F.Zantow, Phys. Rev. {\bf D71}  114510 (2005).}
\label{fig_mdeb}
\end{figure}

The measured~\cite{26sc} spatial string tension in pure glue QCD
corresponds to the values of
$c_\sigma=0.566\pm 0.013$ and  $L_\sigma\equiv \Lambda_\sigma/T_c=0.104\pm 0.009.$
On the Fig.~\ref{fig_mdeb} are shown lattice data \cite{13sc}
and theoretical curves
for the Debye mass in quenched ($N_f=0$) and 2-flavor QCD. Solid lines correspond to our
theoretical prediction, $m_D(T) =c_D \sqrt{\sigma_s(T)}$, with $c_D=2.09$
and for $\sqrt{\sigma_s(T)}$ we exploit the parameters
$c_\sigma=0.564, L_\sigma=0.104$.
The upper solid line is for
the Debye mass in 2-flavor QCD, and the lower -- for quenched QCD. We note that in computing
$m_D(T)$} using (\ref{sigmas}), (\ref{coupl})
all dependence on $N_f$ enters only through the Gell-Mann--Low coefficients $b_0$ and $b_1$.
For comparison we display in the Fig.~\ref{fig_mdeb} dashed lines for  $m_D(T)/T$,
calculated with a perturbative-like ansatz \cite{15sc}.
\be
m_D^{Latt}(T)= A_{N_f} \sqrt{1+\frac{N_f}{6}} g(T) T.
\label{37sc}
\ee

Comparison of our theoretical prediction
with the perturbative-like ansatz shows that both agree
reasonably with lattice data in the temperature interval $T_c < T \leq 5 T_c$;
the agreement is slightly better for our results.
At the same time, in (\ref{37sc}) a fitting constant is used $A_{N_f}\sim 1.5$, which
is necessary even at $T/T_c\sim 5$.
We expect the accuracy of the first approximation of our
approach is around 10\%, taking also into account the bias in the definition
of $\alpha_s^{\rm eff}$ for the gluelump. The temperature region near $T_c$ needs
additional care because i) the behaviour of $\sqrt{\sigma_s(T)}$ deviates from the lattice data
and ii) contribution of chromoelectric fields above
$T_c$  which was neglected above.

\section{Conclusions}

We have studied Debye screening in the hot nonabelian
theory. For that purpose the gauge-invariant definition of
the free energy of the static $Q\bar Q$-pair in the singlet color
state was given in terms of the thermal Wilson loop.
Due to the chromomagnetic confinement persisting at all
temperatures $T$, the hot QCD is essentially nonperturbative. To
account for this fact in a gauge-invariant way the BPTh was
developed for the thermal Wilson loop using path-integral FFS
formalism.  As a result one obtains from the thermal
Wilson loop the screened Coulomb potential with the screening mass
corresponding to the lowest gluelump mass. Applying the
Hamiltonian formalism to the BPTh Green's functions with the
einbein technique the gluelump mass spectrum was obtained. As a
result, we have derived the leading term of the BPTh for
the Debye mass which is the purely nonperturbative,
$$m_D(T)=c_D\sqrt{\sigma_s(T)},~~~~~c_D \approx 2.$$

\section*{Acknowledgments}

The authors are grateful to F. Karsch for supplying us with
numerical data and we thank B.L. Ioffe, A.B. Kaidalov,
D. Kharzeev and V.A. Novikov for useful discussions.
Our gratitude is also to M.A.Trusov for help and useful advices.

The work  is supported by the Federal Program of the Russian
Ministry of Industry, Science, and Technology No.40.052.1.1.1112,
and by the grant  of RFBR No. 06-02-17012, and by the grant for
scientific schools NS-843.2006.2.

\end{document}